\documentclass[runningheads]{llncs}

\usepackage[toc,page]{appendix}

\usepackage{comment}
\usepackage{amsmath}
\usepackage{amsfonts}
\usepackage[T1]{fontenc}
\usepackage[font=scriptsize]{caption}
\usepackage{mwe}

\usepackage{graphicx}
\usepackage{cite}
\newcommand{\MHP}{\ensuremath{\Phi_{\boldsymbol{\ell}}(\boldsymbol{\xi})}}

\newcommand{\PCEW}{\ensuremath{\mathbf{c}_{\boldsymbol{\ell}}(t)}}
\newcommand{\MIi}{\ensuremath{{ \boldsymbol{\ell}}}}
\newcommand{\seed}{\ensuremath{{ \boldsymbol{\xi}}}}

\begin{document}

\title{Polynomial Chaos Expansion Based Nonlinear Filtering of Stochastic Processes}

\author{David Bordenkircher \inst{1}\orcidID{0009-0005-2503-7840} \and
Ruixin Niu \inst{1}\orcidID{0000-0003-2511-9174}}

\authorrunning{D. Bordenkircher and R. Niu.}

\institute{Dept. of ECE, Virginia Commonwealth University, Richmond, VA 23284, USA.\\
\email{\{bordenkirdm,rniu\}@vcu.edu}}

\maketitle              

\begin{abstract}
In Dynamic Data Driven Applications Systems (DDDAS), non-linear continuous-discrete (CD) tracking algorithms have been proposed to recursively estimate stochastic processes which follow continuous-time stochastic differential equations (SDE) using non-linear discrete-time measurement sequences. In this paper, a new filter in this class which is based on the polynomial chaos expansion (PCE) is proposed as an alternative solution to these tracking problems. 
 Using the orthogonality properties of the PCE basis, PCE coefficient-wise prediction and update steps are derived via the Galerkin projection. This differs from previous PCE based filters where collocation points are independently propagated through the motion model and the PCE is recomputed at each time step. For this reason, we call the proposed filter the CD-PCE coefficient filter (CD-PCE-CF). A case study is provided where the CD-PCE-CF and the CD-Extended Kalman Filter (CD-EKF) are used to track a ballistic object undergoing process noise using  radar measurements. It is shown that the proposed method significantly outperforms the CD-EKF in terms of estimation accuracy and stability.
 
\end{abstract}
\section{Introduction}
State estimation is a  critical component of Dynamic Data-Driven Applications Systems (DDDAS). A continuous-discrete (CD) DDDAS system is one in which the target motion follows a continuous time motion law while the measurement sequence is taken in discrete time. The CD model is important since many targets of interest realistically follow continuous motion models. The motion model in a CD system is typically a differential equation (DE) or a stochastic DE (SDE). 
In light of this, CD filtering algorithms became of interest to the tracking community with some of the notable works being  the CD Extended Kalman Filter (CD-EKF) \cite{Jazwinski_book_71}, the CD Unscented Kalman Filter (CD-UKF) \cite{sarkka:2007}, and the CD Cubature Kalman Filter (CD-CKF) \cite{haskin:ieee_tsp_10}.
The fundamental difference in the implementation of a CD filter versus a DD filter is that in the prediction step of a CD filter, one typically uses approximate numerical methods such as Runge-Kutta methods \cite{butcher:1996} in the case of a DE motion law or the Euler-Maruyama or Milstein method \cite{mao:2007,milstein:1975} for a SDE motion law to propagate the state estimate to the next time step.

The polynomial chaos expansion (PCE) \cite{wiener:1938,Cameron&Martin:annal_math_47} provides a means of expressing a random process as a sum of weighted orthogonal basis functions. Recently, the PCE has received attention in the context of non-linear CD tracking thanks to the fact that the PCE representation of a random process enables computationally efficient propagation of the process and all of its statistics without obeying a particular prescribed distribution. Additionally, the properties of the PCE enable multivariate DEs to be decoupled into multiple univariate DEs via methods like the Galerkin projection. In \cite{sun&etal:fusion_19} the PCE was used in conjunction with the Square Root Ensemble Kalman Filter (SREnKF) to track noiseless ballistic trajectories with uncertainty only in the initial condition. In \cite{Yang&etal} the PCE representation was used to propagate the state and parameter uncertainty while the Schmidt-Kalman filter (SKF) is used to update the state with incoming data.

In this work, we propose a CD-PCE filter which recursively propagates and updates the coefficients of the PCE of the target state directly, which is different from all the existing CD-PCE filters. The coefficient prediction rule is obtained by computing the Galerkin projection of the Milstein/Euler-Maruyama solution of the SDE onto each PCE basis function, and the coefficient update rule is derived by taking the Galerkin projection of the linear minimum mean squared error (LMMSE) estimator onto each basis function. In light of this, we call the proposed filter the CD-PCE Coefficient Filter (CD-PCE-CF). 

\section{Motion and Measurement Models}
Assume the target state $\mathbf{x}(t)\in{R}^{d_x}$ evolves through time $t$ according to the It\=o stochastic differential equation (SDE)
\begin{equation}\label{eq:SDE}
    d\mathbf{x}(t)=\mathbf{f}(\mathbf{x}(t),t)dt+\mathbf{G}(\mathbf{x}(t),t)d\mathbf{w}(t)
\end{equation}
where $\mathbf{f}$ and $\mathbf{G}$ are the deterministic state transition model and process noise diffusion function, respectively, $d\mathbf{w}(t)$ is the differential of the Brownian process, and the initial condition $\mathbf{x}(0)$ is distributed as $\mathcal{N}(\mathbf{x}_0,\mathbf{P}_0)$. Suppose we take discrete time measurements $\mathbf{z}(t_k)\in R^{d_z}$ of the target process at intervals indexed by $t_k=kT_s$, where $T_s$ is the sampling period, according to the measurement model
\begin{equation}
    \mathbf{z}(t_k)=\mathbf{h}_{t_k}(\mathbf{x}(t_k))+\mathbf{v}(t_k)
\end{equation}
where $\mathbf{h}_{t_k}$ is a non-linear mapping from $R^{d_x}$ to $R^{d_z}$ and $\mathbf{v}(t_k)\sim\mathcal{N}(\mathbf{\mathbf{0}},\mathbf{R}(t_k))$.
\subsection{Milstein Method Solution}
An approximate solution to the It\=o SDE in (\ref{eq:SDE}) can be computed on the time interval $[t_{k-1},t_k]$, that is, in between measurements, via the Milstein method \cite{milstein:1975} wherein the interval is uniformly partitioned into $K$ sub-intervals with width $\Delta \tau$ indexed by ``virtual time'' $\tau_i$ for $i=1,\ldots,K$ such that $\tau_i-\tau_{i-1}=\Delta\tau$, $\tau_0=t_{k-1}$, $\tau_K=t_{k}$, and the state trajectory is realized by recursively computing
\begin{align}
    \mathbf{x}(\tau_i)=\mathbf{x}(\tau_{i-1})+\mathbf{f}(\mathbf{x}(\tau_{i-1}),\tau_{i-1})\Delta \tau+\mathbf{G}(\mathbf{x}(\tau_{i-1}),\tau_{i-1})\Delta \mathbf{w}(\tau_{i-1})+HOT
\end{align}
where $\Delta\mathbf{w}(\tau_{i-1})=\mathbf{w}(\tau_i)-\mathbf{w}(\tau_{i-1})$ with distribution $\Delta\mathbf{w}(\tau_{i-1})\sim \mathcal{N}(\mathbf{0},\Delta \tau\mathbf{I})$ and the higher order terms (HOT) exist only if the diffusion function depends on the state. 
Both the weak and strong order of convergence of the Milstein method are $\Delta \tau$.
\section{Polynomial Chaos Expansion}
Let $\xi$ be a random variable, sometimes called the ``random seed'', with arbitrarily shaped probability distribution function (PDF) $p(\xi)$ with support $S$ and let $\phi_{j}(\xi)$ be a family of univariate basis functions with $\phi_0=1$ which are chosen such that for all $i>0$, $\phi_i(\xi)$ is orthogonal to $p(\xi)$, i.e.
\begin{equation}
    \langle\phi_i(\xi)\rangle=\int_S\phi_i(\xi)p(\xi)d\xi=0
\end{equation}
Additionally, each pair of basis functions must be orthogonal to each other
\begin{align}
\langle\phi_i(\xi),\phi_j(\xi)\rangle=\int_S\phi_i(\xi)\phi_j(\xi)p(\xi)d\xi=\langle \phi_i(\xi)^2\rangle\delta_{ij}
\end{align}
where $\delta_{ij}$ is the Kronecker delta which equals $1$ if $i=j$ and $0$ otherwise 
and $\langle\phi_i(\xi)^2\rangle$ is the squared $L^2$ norm of the basis function. Clearly, the choice of basis functions depends on the underlying random seed of the target process. For example, when $\xi$ is standard Gaussian, the probabilist's Hermite polynomials form the orthogonal basis \cite{Cameron&Martin:annal_math_47}. In that case, $\langle\phi_i(\xi)^2\rangle=i!$. A univariate random process which can be expressed as a function of some $\xi$, i.e. $x(t,\xi)$ can be approximated by the $L$th order PCE 
\begin{equation}\label{eq:PCE_uni}
    x(t,\xi)\approx
    \sum_{i=0}^L c_i(t)\phi_i(\xi)
\end{equation}
where each PCE coefficient $c_i(t)$ is computed as 
\begin{align}
    c_i(t)=\int_S\phi_i(\xi)x(t,\xi)p(\xi)d\xi
\end{align}
per the orthogonality property. The PCE is extended to the multivariate case by first defining a multivariate random seed. Let $\boldsymbol{\xi}=[\xi_1,\ldots,\xi_{d_\xi}]$ be a random vector with PDF $p(\boldsymbol{\xi})$ of arbitrary i.i.d. RVs with support $S^{d_\xi}=S\times \ldots\times S$ such that $p(\boldsymbol{\xi})=\prod_ip(\xi_i)$. As before, a family of basis functions is needed. First, the notion of the multi-index is helpful for the construction of the multivariate basis. Let $\MIi=(\ell_1,\ldots ,\ell_{d_x})$ be a multi-index where each index $\ell_i\in\mathbb{N}_0$ and the $L_1$ norm of the multi-index is $||\MIi||=\ell_1+\ldots+\ell_{d_x}$. Let $\MHP$ be a family of multivariate basis functions, where, given some choice of $\MIi$, $\MHP$ is developed as the tensor product of the univariate basis functions
\begin{align}
\MHP=\Phi_{(\ell_1,\ldots,\ell_{d_x})}=\phi_{\ell_1}\ldots\phi_{\ell_{d_x}}=\prod_{i=1}^{d_x}\phi_{\ell_i}(\xi_i)
\end{align}
with the order of $\MHP$ equal to $||\MIi||$. 
Leveraging the fact that $\langle\xi_i,\xi_j\rangle=\langle\xi_i\rangle \langle\xi_j\rangle$ and $\langle\phi_i(\xi),\phi_j(\xi)\rangle=0$ for all $i,j$ such that $i\neq j$, the orthogonality property of the univariate basis is readily extended to the multivariate basis
\begin{equation}
    \langle\MHP,\Phi_{\mathbf{k}}(\boldsymbol{\xi})\rangle=\int_{S^{d_\xi}}\MHP\Phi_{\mathbf{k}}(\boldsymbol{\xi})p(\boldsymbol{\xi})d\boldsymbol{\xi}=\langle\MHP^2\rangle\delta_{\boldsymbol{\ell}\mathbf{k}}\label{eq:ORTHOG}
\end{equation}
where
$\delta_{\boldsymbol{\ell}\mathbf{k}}=\prod_{i=1}^{d_x}\delta_{\ell_i k_i}$.
When the basis functions are Hermite polynomials, $\langle\MHP ^2\rangle=\prod_{i=1}^{d_x}\ell_i!$, which we will abbreviate as $\psi^\MIi$.
Given this basis, the multivariate process $\mathbf{x}(t,\boldsymbol{\xi})$ can be approximated by the $L$th order multivariate PCE
\begin{equation}\label{eq:PCE}
    \mathbf{x}(t,\boldsymbol{\xi})\approx
    \sum_{||\boldsymbol{\ell}||=0}^L \PCEW\MHP
\end{equation}
where $\PCEW$ are deterministic vector-valued PCE coefficients
which are computed by projecting (\ref{eq:PCE}) onto the $\MIi$th basis function $\MHP$ to get
\begin{equation}
    \PCEW=\frac{1}{\psi^\MIi}\int\mathbf{x}(t,\boldsymbol{\xi})\MHP p(\boldsymbol{\xi})d\boldsymbol{\xi}=\frac{1}{\psi^\MIi}\langle\mathbf{x}(t,\seed),\MHP\rangle
\end{equation}
Per the Cameron-Martin Theorem \cite{Cameron&Martin:annal_math_47}, the PCE converges to the stochastic process in the $L_2$ sense as the maximum order $L$ goes to infinity.
In practice, the PCE is truncated to some finite order $L$ and in this work, equivalence will be taken for granted between the process and its truncated PCE representation.
 The mean and covariance of the process are easily computed from the PCE coefficients as
\begin{align}
    \langle\mathbf{x}(t,\boldsymbol{\xi})\rangle=\mathbf{c}_0(t)
    ~~~~~\text{cov}(\mathbf{x}(t,\boldsymbol{\xi}))=\sum_{||\MIi||>0}\psi^\MIi\PCEW\PCEW^T
\end{align}
\section{CD-PCE Coefficient Filter}
In this section we present a novel CD filtering algorithm for stochastic processes
in which the state is recursively estimated by directly updating its PCE coefficients. This differs from other filters in the CD class in that all computations are done in the PCE basis function space directly, avoiding unnecessary reconstruction of the state and recomputation of the PCE on an updated state estimate. We accordingly name the filter the CD-PCE Coefficient Filter.
\subsection{Prediction Step}
Recall the Milstein solution to the It\=o SDE. Since we are concerned with the evolution of the state estimate rather than the true state itself, we replace $\mathbf{x}(\tau_i,\seed)$ with $\hat{\mathbf{x}}(\tau_i|t_{k-1},\seed)$ and we replace $\mathbf{x}(\tau_{i-1},\seed)$ with $\hat{\mathbf{x}}(\tau_{i-1}|t_{k-1},\seed)$, which are the predicted states at virtual times $\tau_i$ and $\tau_{i-1}$ given the data up to $t_{k-1}$, giving the state prediction law
\begin{align}
    \hat{\mathbf{x}}(\tau_i|t_{k-1},\seed)=\hat{\mathbf{x}}(\tau_{i-1}|t_{k-1},\seed)+\mathbf{f}(\hat{\mathbf{x}}(\tau_{i-1}|t_{k-1},\seed))\Delta \tau+\mathbf{G}\Delta \mathbf{w}(\tau_{i-1})
\end{align}
where it is assumed that the diffusion function is a state-invariant matrix, for simplicity. Extending the following results to the state-variant diffusion function case can be done trivially. Now, let us assume that the predicted state estimates at virtual times $\tau_{i-1}$ and $\tau_i$ admit the following PCEs
\begin{align}
    \hat{\mathbf{x}}(\tau_{i-1}|t_{k-1},\seed)=\sum_{||\MIi||}\hat{\mathbf{c}}_\MIi(\tau_{i-1}|t_{k-1})\Phi_\MIi(\seed)
\label{eq:predictedPCE}
    ~~~~\hat{\mathbf{x}}(\tau_i|t_{k-1},\seed)=\sum_{||\MIi||}\hat{\mathbf{c}}_\MIi(\tau_i|t_{k-1})\Phi_\MIi(\seed)
\end{align}
Substituting the predicted state estimate PCEs into the Milstein propagation equation gives the result
\begin{align}
    \sum_{||\MIi||}\hat{\mathbf{c}}_\MIi(\tau_i|t_{k-1})\Phi_\MIi(\seed)&=\sum_{||\MIi||}\hat{\mathbf{c}}_\MIi(\tau_{i-1}|t_{k-1})\Phi_\MIi(\seed)\\
    &+\mathbf{f}(\sum_{||\MIi||}\hat{\mathbf{c}}_\MIi(\tau_{i-1}|t_{k-1})\Phi_\MIi(\seed))\Delta \tau+\mathbf{G}\sqrt{\Delta \tau} \seed\nonumber
\end{align} 
The Galerkin projection of this expression onto the $\mathbf{k}th$ basis function yields
\begin{align}
    \psi^\mathbf{k}\hat{\mathbf{c}}_{\mathbf{k}}(\tau_i|t_{k-1})=\psi^\mathbf{k}\hat{\mathbf{c}}_{\mathbf{k}}(\tau_{i-1}|t_{k-1})+\langle\Phi_\mathbf{k},\mathbf{f}(\cdots)\rangle\Delta \tau +\mathbf{G}\sqrt{\Delta \tau}\langle\Phi_{\mathbf{k}},\seed\rangle
\end{align}
where the argument of the state transition function and the argument of the $\mathbf{k}th$ basis functions are suppressed for clarity. Dividing by the normalizing factor cleanly yields a recursion for directly predicting the $\mathbf{k}th$ PCE coefficient using the knowledge of the motion model and the noise statistics
\begin{align}\label{eq:predictionRule}
    \hat{\mathbf{c}}_{\mathbf{k}}(\tau_i|t_{k-1})=\hat{\mathbf{c}}_{\mathbf{k}}(\tau_{i-1}|t_{k-1})+\frac{\Delta \tau}{\psi^\mathbf{k}}\langle\Phi_\mathbf{k},\mathbf{f}(\cdots)\rangle+\frac{\mathbf{G}\sqrt{\Delta \tau}}{\psi^\mathbf{k}}\langle\Phi_{\mathbf{k}},\seed\rangle
\end{align}
The full predicted state or any of its moments can then be re-constructed from these PCE coefficients. We hypothesize that by directly propagating the PCE coefficients, rather than propagating sample points of the state through the motion model and then recomputing the PCE based on the finite set of propagated samples, as in \cite{sun&etal:fusion_19}, the statistics of the state will be more accurately propagated. 
\subsection{Update Step}
At step $K$ of the Milstein method, $\tau_K=t_k$ and the recursive state prediction is complete. At this point, the state is primed for the measurement update. Despite the lack of Gaussianity in the state estimate, let us consider the LMMSE estimator as an expression for computing the data update
\begin{align}\label{eq:MMSEupdate}
    \hat{\mathbf{x}}(t_k|t_k,\seed)=\hat{\mathbf{x}}(t_k|t_{k-1},\seed)+\mathbf{P}_{\mathbf{xz}}(t_k|t_{k-1})\mathbf{S}^{-1}(t_k|t_{k-1})[\mathbf{z}(t_k)-\hat{\mathbf{z}}(t_k|t_{k-1},\seed)]
\end{align}
where $\hat{\mathbf{z}}(t_k|t_{k-1},\seed)$ is the predicted measurement which can be expressed in terms of the predicted state as $\hat{\mathbf{z}}(t_k|t_{k-1},\seed) = \mathbf{h}(\hat{\mathbf{x}}(t_k|t_{k-1},\seed))$ with mean $\hat{\mathbf{z}}(t_k|t_{k-1})=\langle\hat{\mathbf{z}}(t_k|t_{k-1},\seed)\rangle$, $\mathbf{P_{xz}}(t_k|t_{k-1})$ is the cross-covariance between the predicted state and the predicted measurement processes
\begin{align}
    &\mathbf{P}_{\mathbf{xz}}(t_k|t_{k-1})=\langle \hat{\mathbf{x}}(t_k|t_{k-1},\seed)-\hat{\mathbf{c}}_0(t_k|t_{k-1}),[\hat{\mathbf{z}}(t_k|t_{k-1},\seed)-\hat{\mathbf{z}}(t_k|t_{k-1})]^T\rangle  
\end{align}
and $\mathbf{S}(t_k|t_{k-1})$ is the innovation covariance 
\begin{align}
    \mathbf{S}(t_k|t_{k-1})=\langle \mathbf{z}(t_k)-\hat{\mathbf{z}}(t_k|t_{k-1},\seed),[\mathbf{z}(t_k)-\hat{\mathbf{z}}(t_k|t_{k-1},\seed)]^T\rangle+\mathbf{R}(t_k)
\end{align}
Once again, assuming that the updated state estimate $\hat{\mathbf{x}}(t_k|t_k,\seed)$ admits a PCE
\begin{align}\label{eq:posteriorPCE}
    \hat{\mathbf{x}}(t_k|t_k,\seed)=\sum_{||\MIi||}\hat{\mathbf{c}}_\MIi(t_k|t_k)\MHP
\end{align}
the PCE of the predicted state (\ref{eq:predictedPCE}) when $\tau_i=t_k$ and the posterior state (\ref{eq:posteriorPCE}) can be substituted into the LMMSE update rule to yield
\begin{align}
    \sum_{||\MIi||}\hat{\mathbf{c}}_\MIi(t_k|t_k)\MHP=&\sum_{||\MIi||}\hat{\mathbf{c}}_\MIi(t_k|t_{k-1})\MHP\\&+\mathbf{P}_{\mathbf{xz}}(t_k|t_{k-1})\mathbf{S}^{-1}(t_k|t_{k-1})[\mathbf{z}(t_k)-\hat{\mathbf{z}}(t_k|t_{k-1},\seed)]\nonumber
\end{align}
Computing the Galerkin projection of this equation onto the $\mathbf{k}th$ basis function and dividing by the normalizing coefficient yields
\begin{align}\label{eq:updateRule}
    \hat{\mathbf{c}}_\mathbf{k}(t_k|t_k)
    =\hat{\mathbf{c}}_\mathbf{k}(t_k|t_{k-1})+\frac{1}{\psi^\mathbf{k}}\mathbf{P}_{\mathbf{xz}}(t_k|t_{k-1})\mathbf{S}^{-1}(t_k|t_{k-1})\langle \Phi_\mathbf{k},\mathbf{z}(t_k)-\hat{\mathbf{z}}(t_k|t_{k-1},\seed)\rangle
\end{align}
\begin{remark}
    When $\seed\sim\mathcal{N}(\mathbf{0},\mathbf{I})$, one can approximate the expectation bracket terms $\langle\Phi_{\mathbf{k}},\mathbf{f}(\cdots)\rangle$ and $\langle\Phi_{\mathbf{k}},\seed\rangle$ in (\ref{eq:predictionRule}) and $\mathbf{P_{xz}}(t_k|t_{k-1})$, $\mathbf{S}(t_k|t_{k-1})$, and $\langle\Phi_\mathbf{k},\mathbf{z}(t_k)\allowbreak-\hat{\mathbf{z}}(t_k|t_{k-1},\seed)\rangle$ in (\ref{eq:updateRule}) via quadrature point integration.
    In this work, the unscented transform points \cite{Julier&Uhlmann:ieeep04} are used.
\end{remark}
\section{Case Study: Ballistic Object Tracking}
To assess the efficacy of the proposed CD-PCE-CF we compare its performance to the CD-EKF in a  ballistic object tracking problem. The CD-EKF uses the Euler-Maruyama (equivalent to Milstein in the constant diffusion case) method to propagate the state uncertainty and then uses the standard EKF measurement update.
Let the target state be a  6-dimensional vector containing three position coordinates and their corresponding velocities $\mathbf{x}(t)=[x_1(t),x_2(t),x_3(t),\allowbreak v_1(t),v_2(t),v_3(t)]^T$
where position is in km and velocity is in km/s. Let the deterministic motion law of the state $\mathbf{f}(x(t))$ due to Newton's inverse square-law for gravity be
\begin{align}
    &\mathbf{f}(\mathbf{x}(t))=\left[v_1(t),v_2(t),v_3(t),-\eta \frac{x_1(t)}{d(t)^3},-\eta \frac{x_2(t)}{d(t)^3},-\eta \frac{x_3(t)}{d(t)^3}\right]^T
\end{align}
where $\eta=g_eR_e^2$ is the standard gravitation parameter, $g_e$ is Earth's gravitational constant, $R_e$ is Earth's radius, and $d(t)$ is the  distance from its center. 
We choose the diffusion function to be $\mathbf{G}=\text{diag}(\sigma_1,\sigma_2,\sigma_3,\allowbreak \sigma_4,\sigma_5,\sigma_6)$ where all $\sigma_i=0.06$. From this model, 200 individual 100 second ground truth sample trajectories were generated via the Milstein method using a step size of $\Delta \tau=0.01$.

The deterministic measurement function is $\mathbf{h}(\mathbf{x}(t_k))=[r(t_k),\alpha(t_k),\allowbreak \theta(t_k)]^T$, consisting of the range, azimuth, and elevation of the target and
we assume the measurement noise covariance to be $\mathbf{R}=\text{diag}(\sigma_r^2,\sigma_\alpha^2,\sigma_\theta^2)$ where $\sigma_r=0.8$ km, $\sigma_\alpha=0.14$ mrad, and $\sigma_\alpha=0.1$ mrad with the sampling period $T_s=1$ sec. In both filters, the propagation step size is taken to be $\Delta \tau =T_s$, and the maximum order in the CD-PCE-CF is $L=2$ with the basis functions being probabilist's Hermite polynomials.

\begin{figure}[]
  \centering
  \begin{minipage}[b]{0.495\textwidth}
    \centering
    \includegraphics[width = 1.05\textwidth]{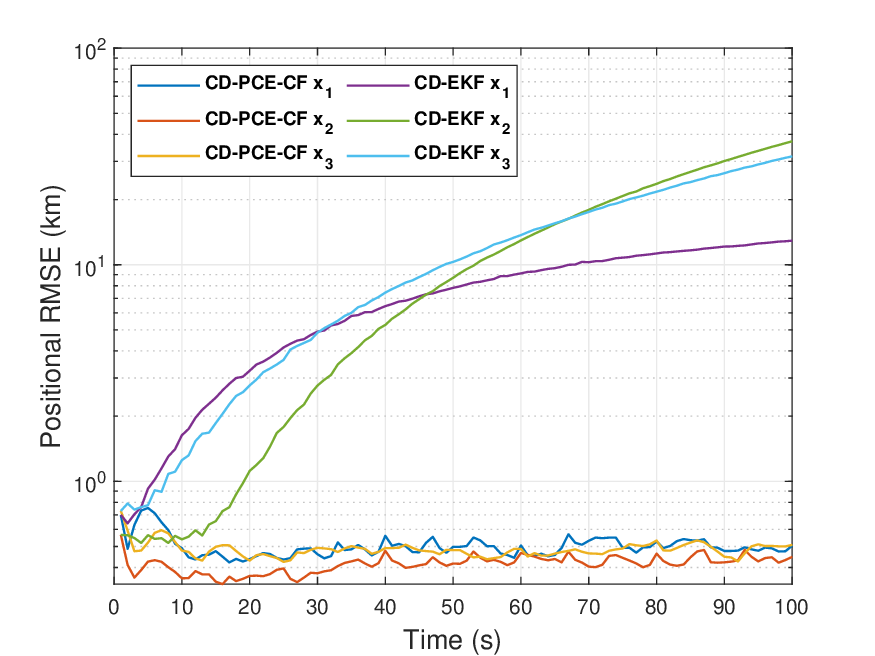}
    \caption{\scriptsize Positional RMSE in each coordinate. }
    \label{fig:RMSE_pos}
  \end{minipage}
  \hfill 
  \begin{minipage}[b]{0.495\textwidth}
    \centering
    \includegraphics[width = 1.05\textwidth]{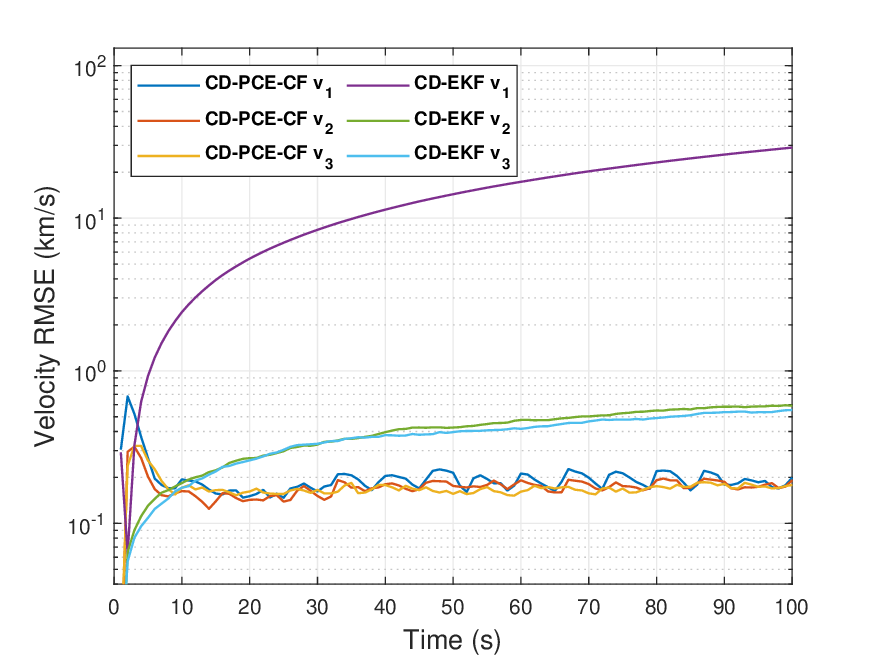}
    \caption{\scriptsize Velocity RMSE in each coordinate.} 
    \label{fig:RMSE_vel}
  \end{minipage}
\end{figure}

Over 200 Monte-Carlo trials, the position and velocity root mean squared errors (RMSE) were computed in each coordinate for both filters and are provided in Figs. \ref{fig:RMSE_pos} and \ref{fig:RMSE_vel}. From Fig. \ref{fig:RMSE_pos}, the positional RMSE of the CD-PCE-CF is consistently on the order of $0.4$ to $0.5$ km with no sign of divergence, while the CD-EKF diverges rapidly over the track duration, nearly reaching an RMSE of $40$ km in $x_2$. Similar results are seen in Fig. \ref{fig:RMSE_vel} where the CD-PCE-CF achieves a nominal velocity RMSE of $0.2$ km/s while the CD-EKF velocity RMSE steadily climbs over the duration. In particular, the CD-EKF poorly estimates the velocity in the $x_1$ axis with an RMSE two orders of magnitude greater than that of the CD-PCE-CF. We deduce that the CD-PCE-CF significantly outperforms the CD-EKF in the ballistic trajectory tracking scenario.
The mean run times for the prediction and update steps of the CD-PCE-CF were determined to be $2.341$ ms and $2.764$ ms, respectively, while the mean run times for the corresponding steps of the CD-EKF were determined to be $0.021$ ms and $0.094$, respectively. Despite increased run-time, we deem the performance trade-off beneficial. 
\section{Conclusion}
A new  CD-PCE Coefficient Filter was proposed  in this work to accurately propagate and update the coefficients of the PCE of a random process directly so as preserve the most statistical information about the process and consequently obtain an accurate state estimate. The prediction and update rules for the PCE coefficients were derived by computing the Galerkin projection of the Milstein state propagation rule and the LMMSE update rule onto each PCE basis function. Numerical results show that for a low maximum order of $L=2$, the proposed filter drastically outperforms the CD-EKF in the noisy ballistic tracking problem. In future work, the CD-PCE-CF will be compared more extensively to competing algorithms and higher order PCEs will be considered. 
\begin{credits}
\subsubsection{\ackname} 
This work was supported in part by the Air Force Office of Scientific
Research under Grant FA9550-25-1-0328, and the Convergence Lab Initiative at Virginia Commonwealth University (\url{https://www.cli-vcu.org}).
\vspace{-0.2in}
\end{credits}

\bibliography{Book,Journal,Conf}

\bibliographystyle{IEEEtran}

\newpage

\end{document}